# Graphical Abstract

## First-principles calculations of structural and bonding properties of Li-doped tetrahedrite thermoelectrics

Krzysztof Kapera,Andrzej Koleżyński

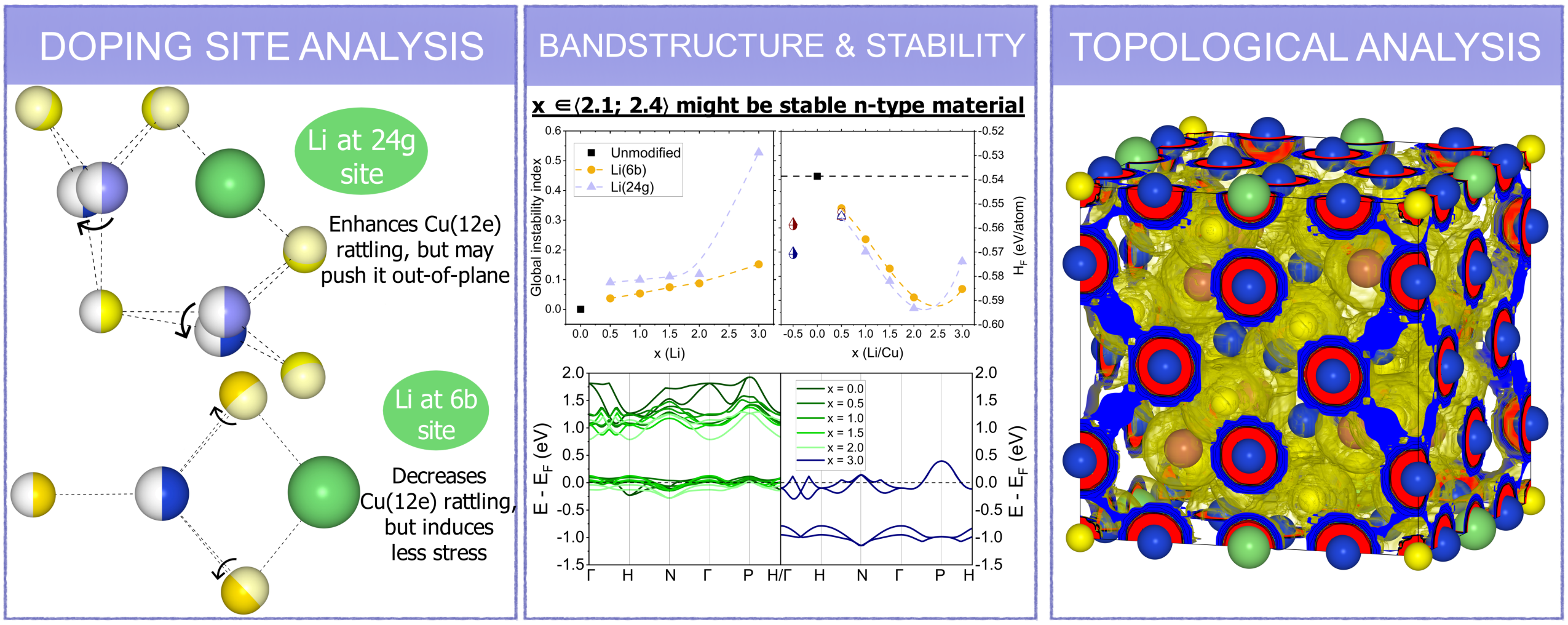

# Highlights

## First-principles calculations of structural and bonding properties of Li-doped tetrahedrite thermoelectrics

Krzysztof Kapera,Andrzej Koleżyński

- Li-doped (into 24g and 6b sites) tetrahedrite were investigated
- Enthalpy of formation and global instability index shows stability drop at $x > 2.0$
- Dopants at 6b and 24g site affect ratting Cu(12e) drastically different
- Band structure calculations predict tetrahedrite to be *n*-type for $x > 2.0$

# First-principles calculations of structural and bonding properties of Li-doped tetrahedrite thermoelectrics

Krzysztof Kapera[a,*], Andrzej Koleżyński[a]

[a]*Department of Silicate Chemistry and Macromolecular Compounds, Faculty of Materials Science and Ceramics, AGH University of Krakow, al. A. Mickiewicza 30, Kraków, 30-059, Poland*

ABSTRACT

Tetrahedrite (copper antimony sulfosalt) is a promising *p*-type thermoelectric material due to its very low intrinsic thermal conductivity and moderately-high power factor, with one of the limitations being the lack of *n*-type variant to create a thermoelectric generator. In this paper, DFT calculations have been carried out to study tetrahedrite doped with Li into structural voids, $Li_xCu_{12}Sb_4S_{13}$ ($0 \leq x \leq 3$). Enthalpies of formation show that the introduction of Li into both 6b and 24g sites is energetically favorable. Dopants in those positions differently affect the rattling Cu(12e) behavior, as well as vary in the magnitude of induced local disorder. Topological analysis of charge density classifies tetrahedrite as a closed-shell, ionic system of interactions with some degree of covalency. The addition of Li increases bond strain and decreases structural stability. Electronic band structure shows that for $x > 2.0$, material becomes *n*-type; however, results are not precisely conclusive on whether structure will be synthesizable, which should be determined experimentally.

## 1. Introduction

Thermoelectric materials have garnered significant attention in the realm of materials science and energy research due to their unique ability to convert waste heat into usable electrical power. This phenomenon relies on the Seebeck effect, where a temperature gradient across a material induces an electric voltage. The performance of a thermoelectric material is quantified by the dimensionless figure of merit, $ZT = S^2\sigma T/\kappa$, where $S$ $[VK^{-1}]$ is the Seebeck coefficient, $\sigma$ $[Sm^{-1}]$ is electrical conductivity, $T$ $[K]$ is absolute temperature and $\kappa$ $[Wm^{-1}K^{-1}]$ is total thermal conductivity (composed of lattice $\kappa_L$ and electronic $\kappa_E$ part). Thus, an ideal thermoelectric material possesses high Seebeck coefficient (to guarantee high voltage), high electrical conductivity (which minimizes Joule heating), and low thermal conductivity (which ensures the temperature gradient is high).

However, optimization of $ZT$ is challenging – increasing electrical conductivity generally increases $\kappa_E$, as stated in the Wiedemann-Franz law[32] (expressed by formula $\kappa_E/\sigma = LT$, where $L$ is Lorenz number), making it difficult to enhance $\sigma$ without affecting thermal conductivity. Moreover, simultaneous increase in $S$ and $\sigma$ is limited. As the Pisarenko relation[35] states:

$$S = \frac{8\pi^2 k_B^2}{3eh^2} m^* T \left(\frac{\pi}{3n}\right)^{2/3} \quad (1)$$

where $k_B$ is Boltzmann constant, $h$ is the Planck constant, $e$ is elementary charge, $m^*$ is effective mass and $n$ is carrier concentration. Seebeck coefficient is inversely proportional to $n$ and directly proportional to $m^*$, while electrical conductivity exhibits opposite dependencies ($\sigma = ne\mu = \frac{ne^2\tau}{m^*}$, where $\mu$ is carrier mobility and $\tau$ is relaxation time).

Band engineering through doping is a popular strategy to optimize carrier concentration and their mobility, thereby increasing electrical conductivity. Even though, according to eq. 1, the Seebeck coefficient is inversely proportional to carrier concentration, increase in $n$ leads to band degeneration and an increase in effective mass $m^*$[8]:

$$n_H = \frac{1}{eR_H} = A^{-1}\frac{N_v(2m_b^* k_B T)^{3/2}}{3\pi^2 h^3} F_0^{3/2} \quad (2)$$

where $N_v$ is band degeneracy and $F_0$ is Fermi-Dirac integral. A local increase in the density of states $g(E)$ in narrow energy range close to Fermi level is also a valid strategy. According to Mott's formula[11]:

$$S = \frac{\pi^2}{3}\frac{k_B}{q} k_B T \left\{\frac{d[ln(\sigma(E))]}{dE}\right\}_{E=E_F} \quad (3)$$

which means that the Seebeck coefficient depends on derivative of electrical conductivity $\sigma(E) = n(E)q\mu(E)$ at the Fermi level. Increased energy-dependence of $n(E)$ via a local increase of $g(E)$ results in higher Seebeck coefficient.

Tetrahedrite, $Cu_{12}Sb_4S_{13}$, is a p-type conductor studied for its application as a thermoelectric material at medium to high temperatures ($500 - 700K$), motivated by several factors. It has a complex crystal structure with 58 atoms per unit cell (a.p.f.u.), high symmetry, and intrinsically low thermal conductivity ($1.0 - 1.6$ $Wm^{-1}K^{-1}$ for pure, synthetic samples[6]); however, significantly lower values have been achieved in various studies[27, 14, 18]. The Seebeck coefficient is positive, indicating that the majority carriers are holes, with values ranging from 100 to 150 $\mu VK^{-1}$. Tetrahedrite has metallic character, with typical $\sigma$ values on the order of $1 - 4 \cdot 10^4$ $Sm^{-1}$. Several research teams

*Corresponding author

kaperak@agh.edu.pl (K. Kapera); kolezyn@agh.edu.pl (A. Koleżyński)

ORCID(s): 0000-0003-3388-3885 (K. Kapera); 0000-0002-4682-8300 (A. Koleżyński)

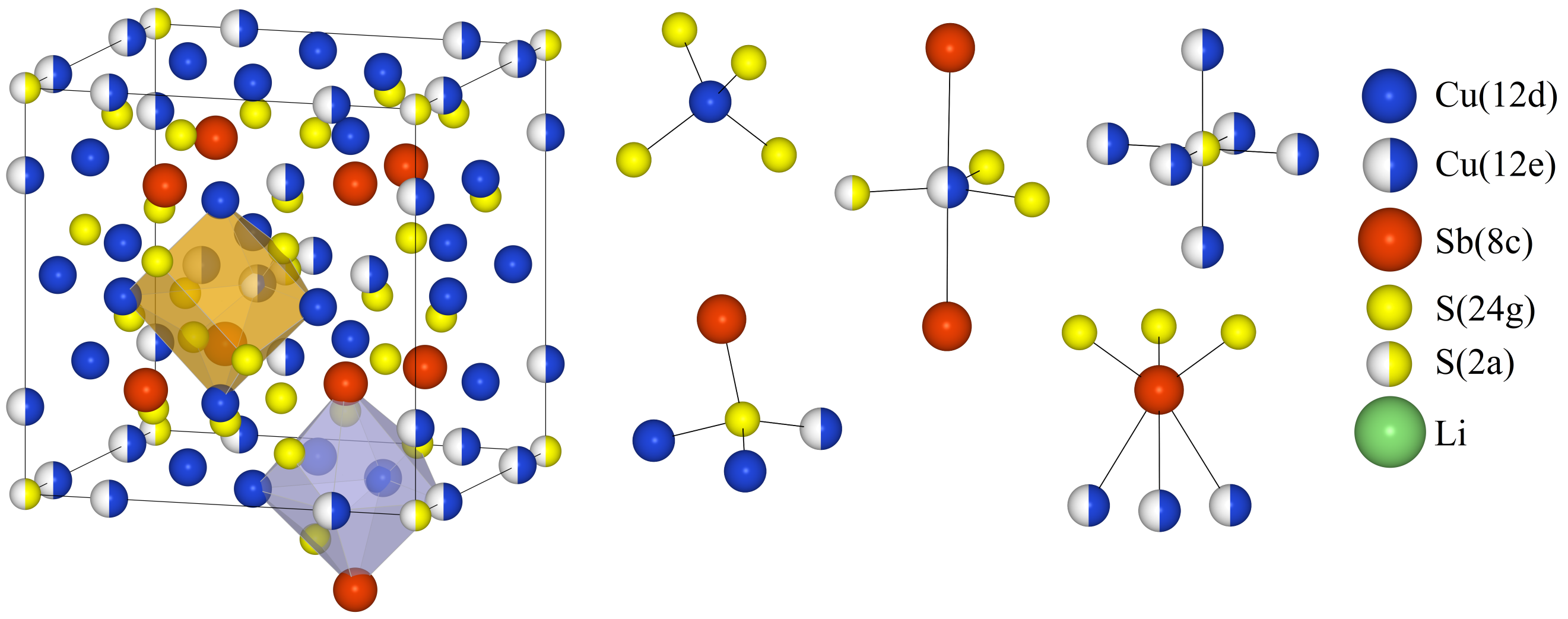

**Figure 1:** Tetrahedrite's unit cell with 58 atoms, along with structural units. Polyhedra mark structural voids with Wyckoff position 6b (orange polyhedron) and 24g (teal polyhedron). Li atom in the legend is for a future reference.

have succeeded in reaching $ZT \sim 1$ for doped tetrahedrite [6, 18, 12, 19, 10, 21, 17, 28, 34]. Furthermore, naturally occurring tetrahedrite is the most abundant sulfosalt on Earth, and a processing method for its use as a thermoelectric material has been demonstrated [18].

Wuensch first described tetrahedrite's structure[33]. It crystallizes in a cubic crystal system (space group $I\bar{4}3m$), with atoms occupying five different Wyckoff positions: Cu(12d), Cu(12e), Sb(8c), S(24g), and S(2a). Tetrahedrite's unit cell and structural units can be seen in **Figure 1**. Stoichiometrically, the tetrahedrite's formula can be then written as $Cu(12d)_6Cu(12e)_6Sb(8c)_4S(24g)_{12}S(2a)$. Out of six Cu(12d) cations, four possess charge $Cu^+$ and two are $Cu^{2+}$, with a random distribution (known as valency-imposed double-site occupancy). Tetrahedrite meets Slack's phonon-glass and electron-crystal (PGEC) criteria for thermoelectric materials[25]. Strong, anharmonic, low-frequency, and high-amplitude out-of-plane vibrations of Cu(12e) atoms are responsible for scattering phonons and significantly lowering tetrahedrite's thermal conductivity. The atom displacement parameter (ADP) of $U_\perp = 0.13$ Å [26] indicates that Cu(12e) behaves similarly to rattling atoms in cage-like structures (such as clathrates), even without a traditional cage. Studies concerning this phenomenon have been done by a few teams. Lai et al.[15] carried out *ab initio* simulations and synchrotron diffraction measurements. According to their results, most of Cu(12e) atoms are located on the $S_3$ plane; however, the out-of-plane distance can be as large as 0.8 Å (with the average distance being 0.27 Å and a standard deviation of 0.18 Å). Cu(12e) approaching Sb(8c) increases the bond order from 0.1 to $\sim 0.35$. Simulations indicated that Cu(12e) can create bonds with both Sb(8c) atoms, but only one at a time. According to the authors, Cu(12e) behavior is due to its interaction with antimony lone electron pairs. Umeo et al.[29] studied the effect of pressure on rattling in tetrahedrite and type-I clathrates by means of specific heat and synchrotron diffraction measurements. Their results suggest that Cu(12e) vibrations are due to "chemical pressure," i.e., forces acting upon Cu(12e) coming from interactions with nearby S atoms, which naturally increase as the distance between Cu(12e) and S shortens. Above a pressure of 2.4 GPa, anharmonic rattling of Cu(12e) ceases, and it moves to a new 24g position, out of the $S_3$ plane. The role of antimony in Cu(12e) rattling is described as minor. Suekuni et al.[26] studied crystal structure and phonon dynamics in the tetrahedrite family by means of synchrotron diffraction and inelastic neutron scattering experiments. Any relation between Cu(12e) and Sb/As(8c) distance has not been found; however, they established that rattling energy is inversely proportional to ADP. The authors suggested that "chemical pressure" (inversely proportional to the area of the $S_3$ triangle) is the main force driving Cu(12e) rattling.

There have been many attempts to improve tetrahedrite's thermoelectric properties via doping, primarily by introducing *d*-block metals into copper's sublattice. Some studies experimented with substituting Sb/S or adding elements into structural voids. Most transition metals introduced into copper's sublattice occupy $Cu^+$(12d) Wyckoff position, however sometimes it is energetically favorable for $Cu^{2+}$(12d) to change to $Cu^+$(12d) and substitution for both Cu(12d) and Cu(12e) are possible. Vaqueiro et al.[30], found that in a Cu-rich structure, there was another partially occupied Cu(24g) site (0.28, 0.28, 0.041). There has been only one study that successfully produced n-type tetrahedrite via Fe-doping in the low $80 - 310$K temperature range ($Cu_{10.5}Fe_{1.5}Sb_4S_{13}$)[31].

In this paper, first principle DFT calculations have been used to investigate tetrahedrite doped with Li into structural voids with Wyckoff positions 6b and 24g ($Li_xCu_{12}Sb_4S_{13}$, where $x = 0.5, 1.0, 1.5, 2.0, 3.0$). The reasoning behind lithium is as follows: alkali metals (such as Li) have not been previously studied as a dopant in tetrahedrite; Li is a strongly electropositive atom, which should introduce ionic bonds into the system, reduce thermal conductivity, and potentially enable doping into structural voids to achieve *n*-type material. DFT calculations performed using the WIEN2k computational package have provided insights into site preferences and enthalpies of formation. Topological analysis using Critic2 software has facilitated investigation into chemical bonds, local disorder, and charge transfer. Brown's Bond Valence Model was used to calculate global instability and bond strain indices. Changes in Cu(12e) behavior and the electronic band structure have been examined

## 2. Theoretical approach and computational details

*Ab initio* calculations were carried out using WIEN2k 19.1 computational package[2, 4], which employs Full-Potential Linearized Augmented Plane Wave (FP-LAPW) approximation[3, 24], within Density Functional Theory (DFT) formalism. Output data for relaxed structures from WIEN2k were used as input data for CRITIC2 software [22], which allows for topological analysis of total charge density in accordance with Bader's Quantum Theory of Atoms in Molecules (QTAiM)[1]. Bond lengths and valencies were used to calculate global instability and bond strain indices using Brown's Bond Valence Model (BVM) equations [5]

For WIEN2k calculations, following parameters were used: GGA PBEsol exchange-correlation functional, muffin-tin radii $R^{S}_{MT} = 1.8$, $R^{Cu}_{MT} = 2.1$, $R^{Sb}_{MT} = 2.3$, $R^{Li}_{MT} = 1.9$, $256 - k$ points in the irreducible Brillouin Zone for the undoped tetrahedrite structure and a matching number of points (10x10x10 grid) for modified structures, $R_{KMax} = 8.0$, $LM = 8.0$, $G_{Max} = 20$, convergence criteria for energy $\Delta E = 10^{-5}$ Ry, charge $\Delta q = 10^{-5}$ e and force $\Delta F = 10^{-1}$ mRy $a_0^{-1}$.

To thoroughly investigate how structural modifications affect tetrahedrite's properties and how different arrangements of atoms influence the enthalpy of formation, the following structures were created:

- Undoped tetrahedrite $Cu_{12}Sb_4S_{13}$ structure
- Doped tetrahedrite in which Li was introduced in either structural void 6b $(0, \frac{1}{2}, \frac{1}{2})$ or 24g $(x, x, z)$, $Li_x$(6b/24g)$Cu_{12}Sb_4S_{13}$ $(x = 0.5, 1.0, 1.5, 2.0, 3.0)$
- Doped tetrahedrite in which Li was introduced in either structural void 6b or 24g, with Cu(12e/12d) deficiency $Li_{0.5}$(6b/24g)$Cu_{11.5}Sb_4S_{13}$
- Doped tetrahedrite in which Li was introduced in position 12e or 12d, with Cu occupying structural void 6b or 24g, $Li_{0.5}$(12e/12d)$Cu_{0.5}$(6b/24g)$Cu_{11.5}Sb_4S_{13}$

All structures were optimized; lattice parameters and atomic positions were allowed to change with respect to total energy and forces acting upon. P1 symmetry was used for all structures in order to not constrain atoms during relaxation, which is more computationally demanding, however yields better quality of results.

## 3. Results and discussion

After optimizing internal parameters for all structures and obtaining their total energies, enthalpy of formation, $H_F$ has been calculated. It is defined as the energy difference between total energy of a given structure and the sum of total energies of pure elements in their thermodynamically stable structures (for which calculations, including relaxation also has been performed). For example, $H_F$ for $LiCu_{12}Sb_4S_{13}$:

$$H_F^{LiCu_{12}Sb_4S_{13}} = E_{tot}^{LiCu_{12}Sb_4S_{13}} - 12 \cdot E_{tot}^{Cu} - 4 \cdot E_{tot}^{Sb} - 13 \cdot E_{tot}^{S} - E_{tot}^{Li} \quad (4)$$

where $H_F$ and $E_{tot}$ are in eV/atom.

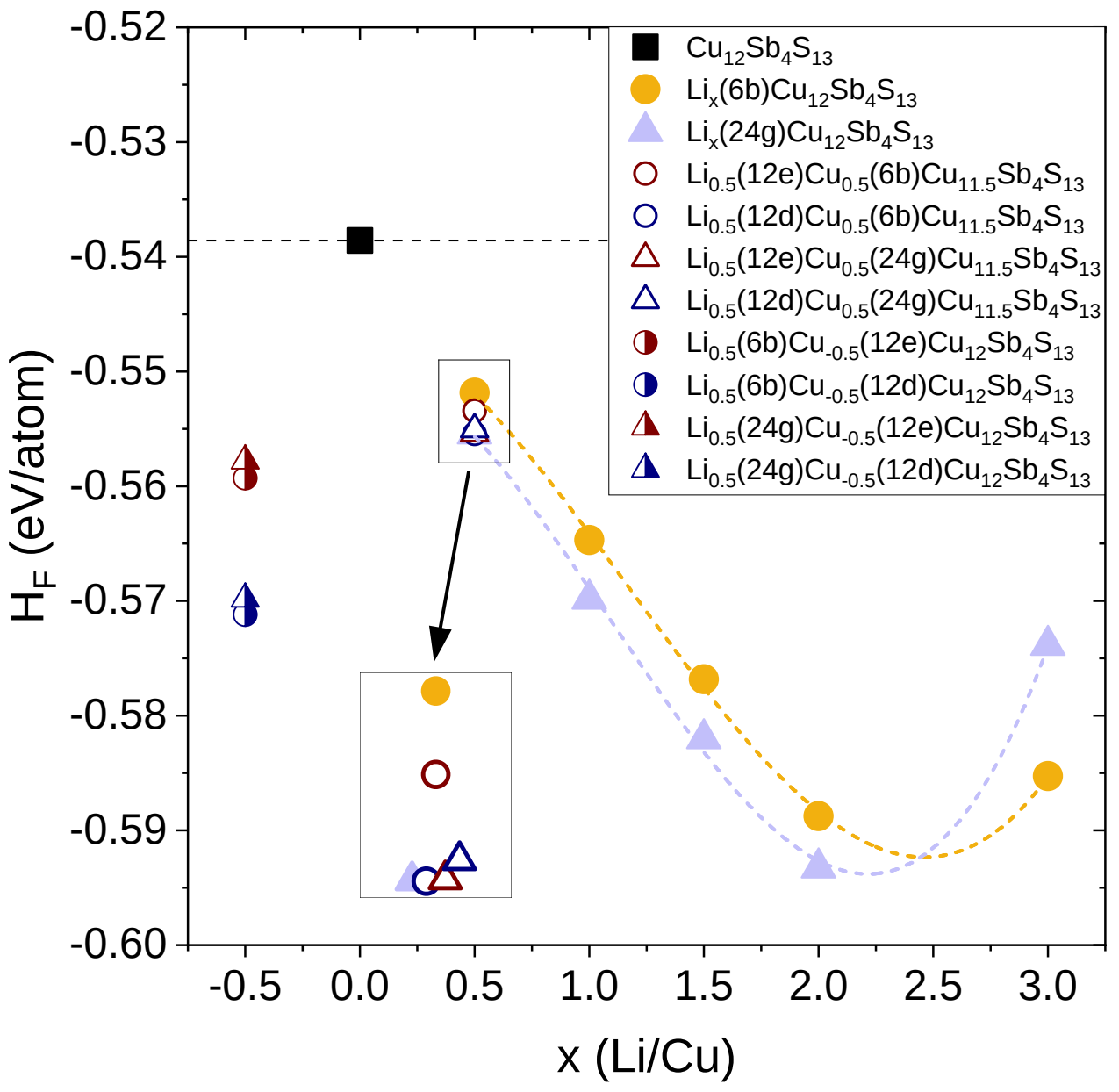

**Figure 2:** Enthalpies of formation for optimized and relaxed structures, calculated based on the Equation (4). The fitted polynomial is there mostly to guide the eye, as it does not represent any real physical relation.

Results of enthalpy of formation (**Figure 2**) provides a few key points. For one, calculations predict that Li atoms may be incorporated into the structural voids, with 24g Wyckoff position being moderately more stable than 6b. The enthalpy of formation for Li-doped structures decreases with Li content, then increases for $x = 3.0$, suggesting a solubility limit in the range $2.0 \leq x < 3.0$. For $x = 0.5$, a plethora of structures have similar $H_F$ values between $-0.55$ and $-0.56$ eV/atom, suggesting a high probability of different defects being present simultaneously in Li-doped tetrahedrite. Finally, structures with Cu-deficiency ($x = -0.5$ on Figure 2) seems to be more energetically stable with a Cu(12d) vacancy, which is consistent with multiple findings by other groups (mostly done for transition metals as dopants).

As expected, unit cell lattice parameter, shown in **Figure 3** increases (approximately) linearly with Li content. Fitting yields formulae $0.0451x$ Å and $0.0522x$ Å for Li(6b) and Li(24g) respectively, where $x$ represents atoms per formula unit (a.p.f.u) in $Li_x(6b/24g)Cu_{12}Sb_4S_{13}$. This is roughly half of what has been reported by Levinsky et al. for Mg[16] (0.1 Å increase per Mg a.p.f.u.). This discrepancy is not explained by the difference in ionic radii (59 pm for Li and 57 pm for Mg[23]). Based on our previous *ab initio* results for Mg-doped tetrahedrite[13] we notice that topological volume (calculated by CRITIC2) of Li is much smaller than topological volume of Mg (20 and 35 $au^3$ respectively). This probably means that relying solely on ionic radius might not be sufficient to make claims about lattice parameter in tetrahedrite.

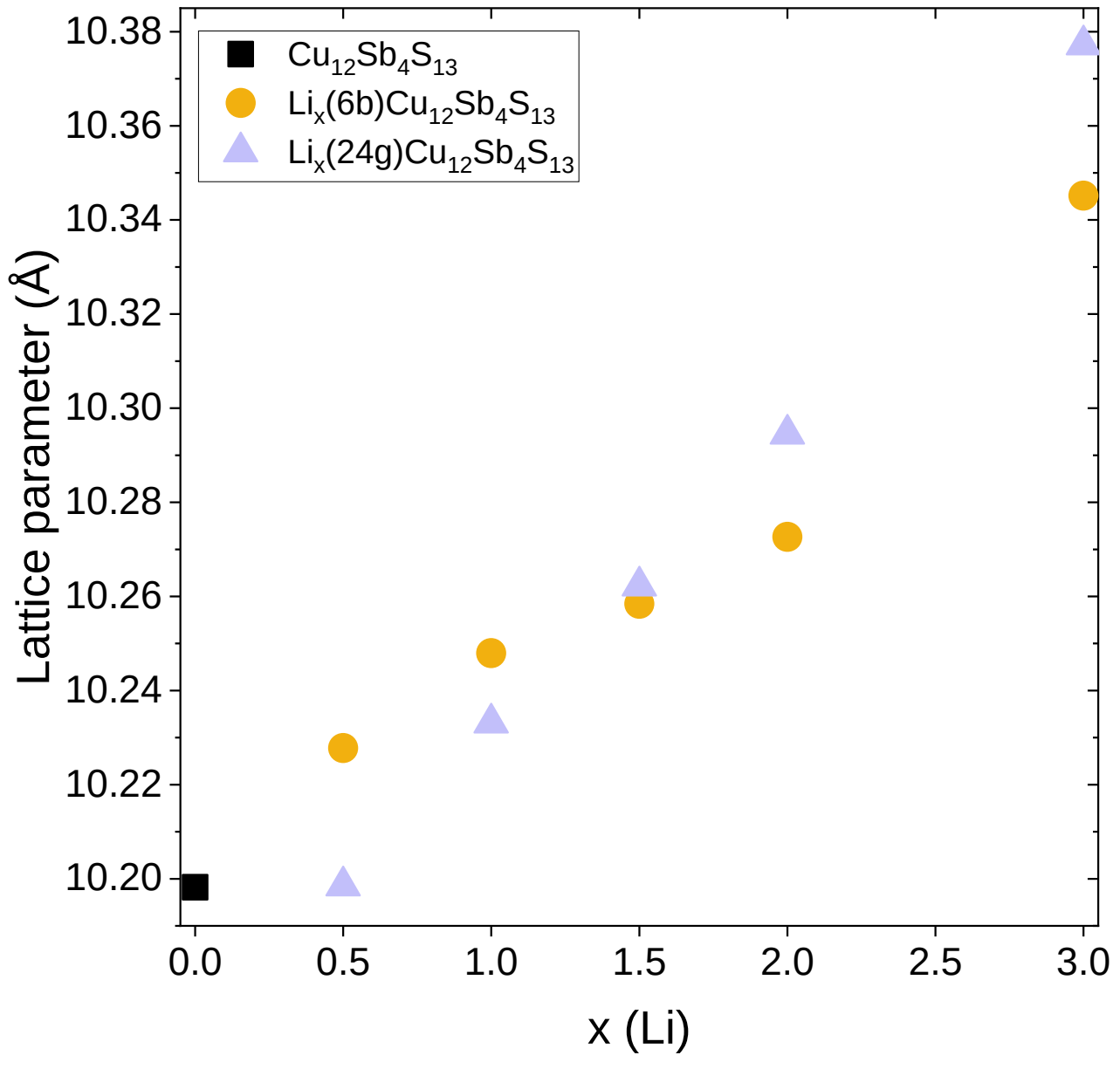

**Figure 3:** Lattice parameter for optimized Li-doped tetrahedrite structures obtained with WIEN2k calculations.

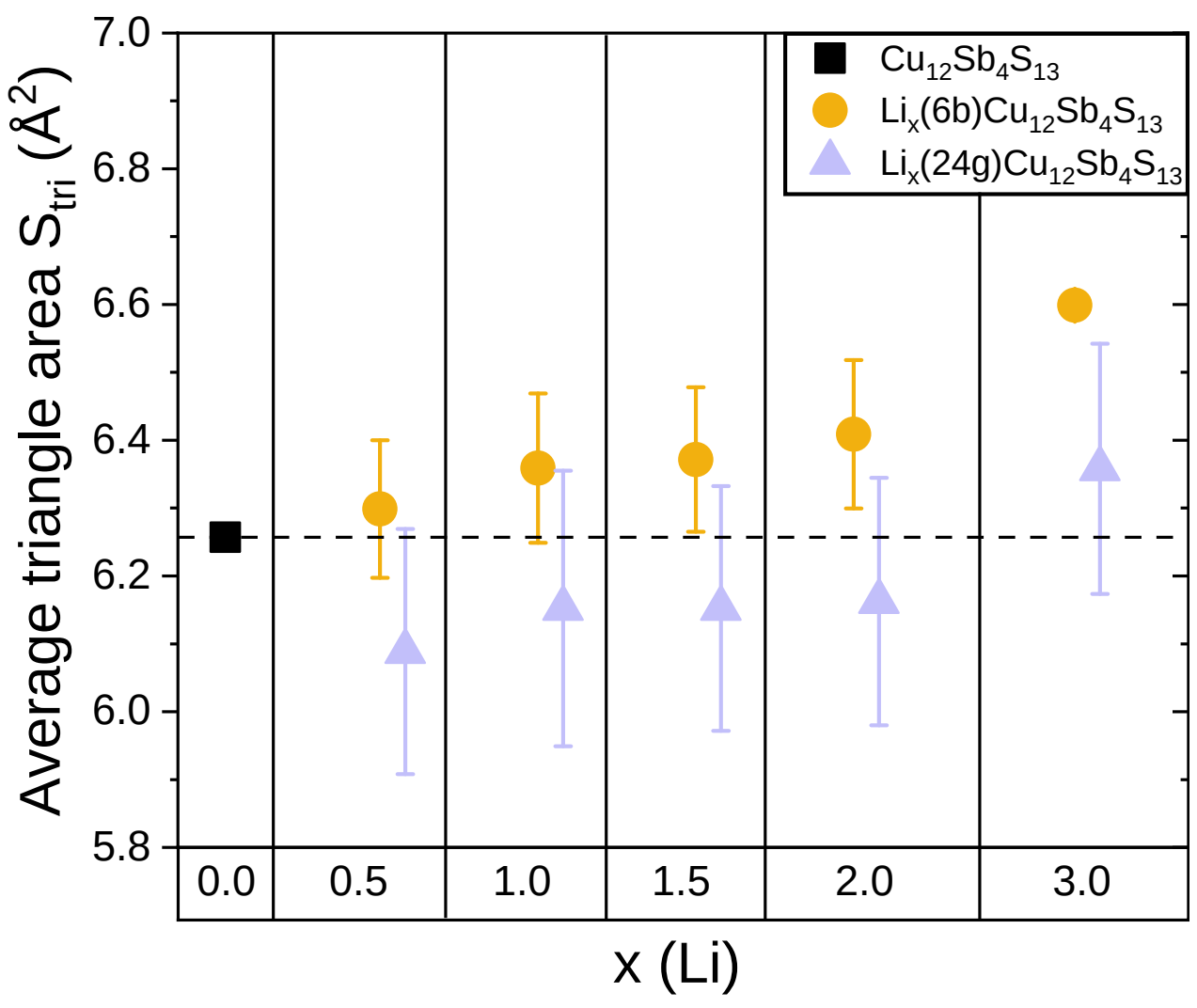

**Figure 4:** Averaged area of the $S_{tri}$ triangle around Cu(12e) sites. Error bars indicate standard deviation.

As mentioned earlier, area of the $S_3$ triangle, $S_{tri}$ is a crucial parameter that is at least partially responsible for Cu(12e) rattling, with smaller area resulting in higher "chemical pressure" exerted and higher ADP. Calculated average $S_{tri}$ areas for all structures are presented in **Figure 4**

There is a clear distinction between structures, with the average area $S_{tri}$ increasing for Li at the 6b site and decreasing for Li at the 24g site. This is due to differences in local environment of the dopants at 6b and 24g sites, which will be explained in detail further into the text. Based on linear relation by Suekani et al.[26], we estimate that any rattling would cease for $S_{tri} = 6.55$ Å$^2$, which applies to Li(6b) structure with $x = 3.0$. In contrast, Li(24g) structures should have enhanced rattling, with ADP up to $U_\perp = 0.27$ Å for $x = 0.5$, which is double the ADP of undoped tetrahedrite ($U_\perp = 0.13$ Å). Umeo et al.[29] studied effect of pressure on rattling in $Cu_{10}Zn_2Sb_4S_{13}$ (CZSS), a structure very similar to our tetrahedrite, and their results are consistent with ours. Umeo et al. obtained $S_{tri} = 6.07$ Å$^2$ for CZSS under 1.9 GPa pressure, which is almost identical to our $Li_{0.5}(24g)Cu_{12}Sb_4S_{13}$ ($S_{tri} = 6.088$ Å$^2$). The authors speculated that above 2 GPa, the Cu(12e) atom is forced out of the $S_3$ plane into a new position where it strongly bonds with antimony, thereby ceasing its rattling. We have observed that some Cu(12e) atoms in the calculated Li(24g) structures are indeed out-of-plane. Mostly, however, they should remain in the range of such chemical pressure acting upon Cu(12e) that would decrease rattling energy and increase ADP, consequently increasing phonon scattering and lowering thermal conductivity. Additionally, DFT calculations are done for $0K$ temperature. At higher temperatures and in real tetrahedrite, thermal expansion should help in ensuring proper rattling.

To understand how lithium atoms could change the behavior of Cu(12e), the local impact of dopants on nearest neighbors was investigated (see **Figure 5**)

Structural void 6b is surrounded by symmetrically distributed four S(24g), four Cu(12d) and two Cu(12e) atoms. As can be seen at Figure 5a, only distance to S(24g) increases by up to 6%. This change means that area of $S_3$ triangle increases (see Figure 5c) due to the triangle being more equilateral. A different picture is painted for Li at structural void 24g - here dopant is surrounded by asymmetrically distributed two Sb(8c), two Cu(12d), two Cu(12e) and three S(24g) atoms. The introduction of Li into the 24g void strongly affects the antimony positions, with one of the Sb atoms moving away from Li by up to 20% of its original distance (the other Sb atom moves away by only 5%, further increasing asymmetry). For higher Li concentrations, S(24g) and Cu(12d) move closer to the Li atom, which should strengthen their bond to Li. Finally, Cu(12e) atom moves away from the Li dopant, in the direction perpendicular to $S_3$ triangle and out of the plane, similar to what has been reported in other works. It is worth mentioning that for Li(24g) structures, not all Cu(12e) atoms were pushed out of $S_3$ plane.

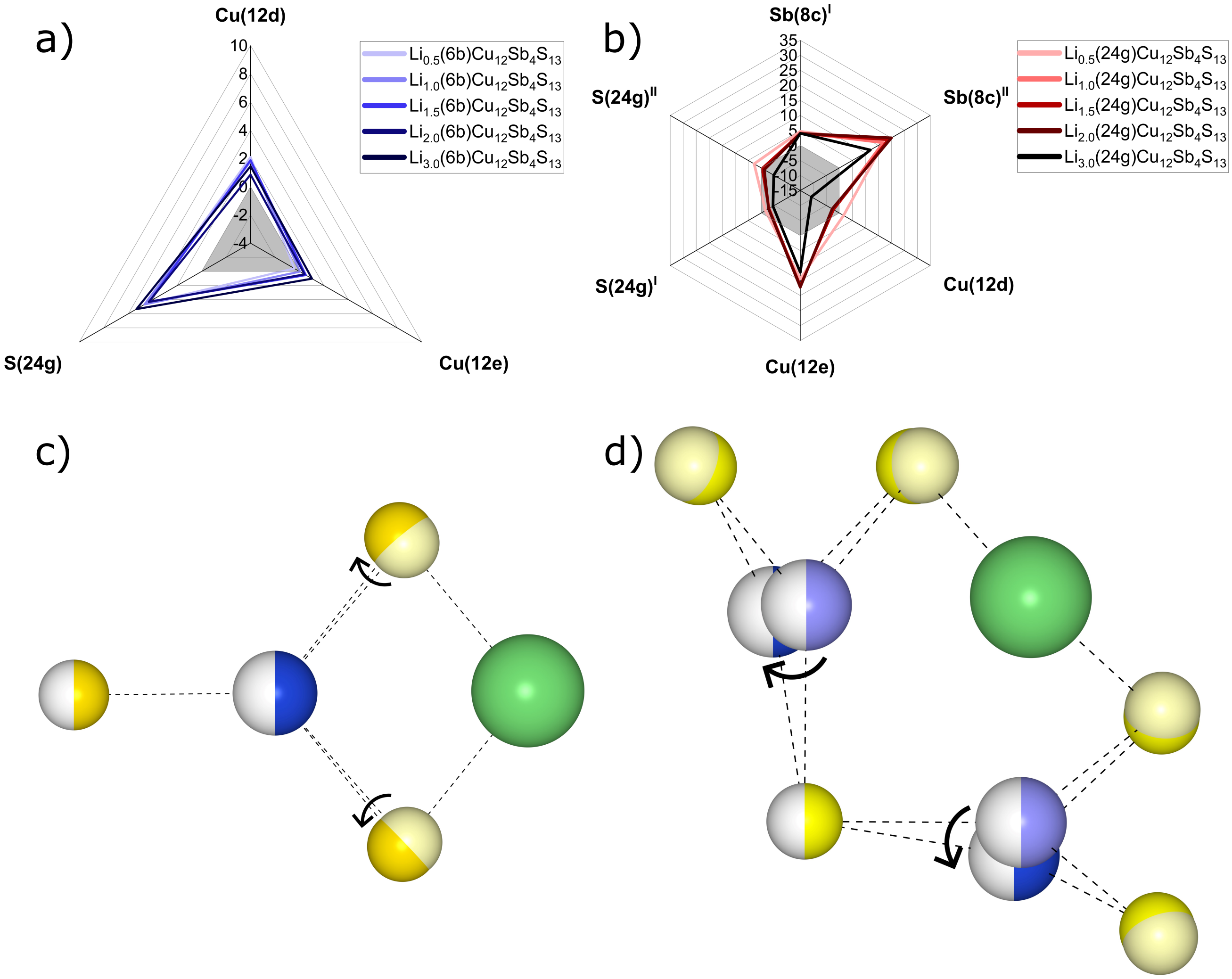

**Figure 5:** Upper part: % distance change between center of **a) Li(6b)**, **b) Li(24g)** and neighboring atoms. Lower part: Illustrative representation of change in atomic positions in local environment of **c) Li(6b)**, **d) Li(24g)** after introducing it into the structure. More saturated colors denote atomic position after Li has been introduced, with arrows marking their motion.

| Structure name | Triangle area, $S_{tri}$ [Å$^2$] | Distance Cu(12e) – Sb(8c) [Å] | Local presence of dopant |
|---|---|---|---|
| *Undoped* | 6.26 | 3.23 | – |
| *Li(24g) v1* | 6.48 | 3.24 | – |
| *Li(6b)* | 6.47 | 3.31 | Yes |
| *Li(24g) v2* | 6.04 | 2.71; 4.16 | Yes |

**Table 1**
Structure details for which additional calculations were made with Cu(12e) displaced out of the equilibrium position

Local disorder affecting Cu(12e) rattling also has been investigated through additional calculations aiming to capture freeze-frames during its motion. Four different structures with Cu(12e) displaced perpendicularly out of the $S_3$ plane have been created to study changes in the total energy of the system and forces acting upon Cu(12e). Details can be seen in **Table 1**. In structures *Undoped, Li(24g) v1* and *Li(6b)*, Cu(12e) atom is on the $S_3$ plane, while in the *Li(24g) v2* structure, the Cu(12e) atom is out-of-plane at equilibrium.

For the *Undoped* structure, rattling seems to be energetically favorable since total energy of the system decreases (see **Figure 7**). Forces acting on Cu(12e) increase til $\sim$ 1.1 mRy au$^{-1}$ and then slightly decrease. For 0.36 Å displacement, vector forces suddenly changes direction, away from antimony atom (see appendix, Figure 11a). The sharp drop in energy for *Undoped* at 0.36 Å might be due to increased bond order with antimony.

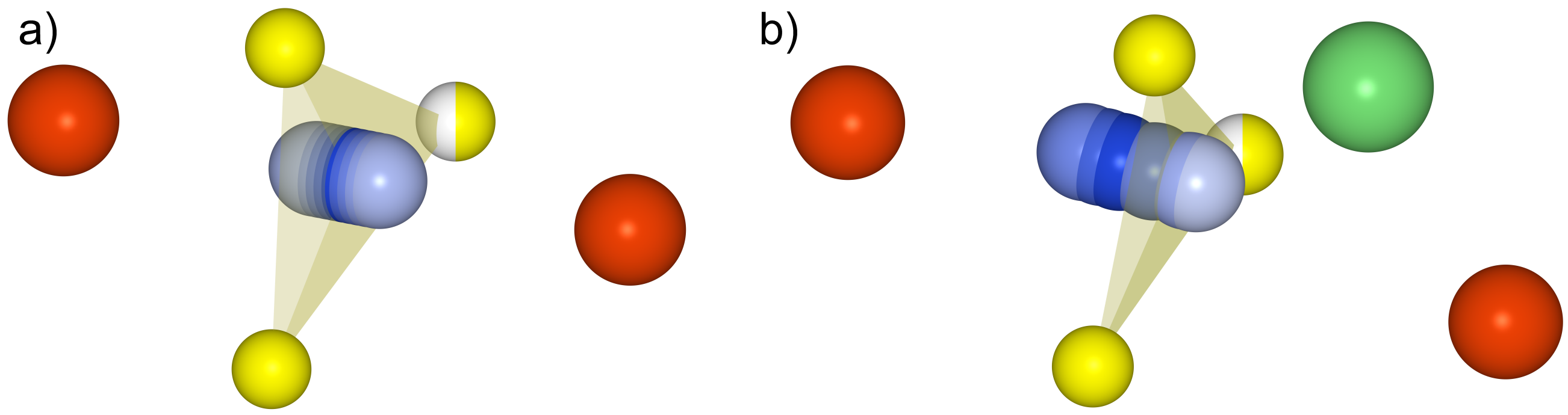

**Figure 6:** Positions of displaced Cu(12e) atom. The darkest blue hue denotes initial, equilibrium position of Cu(12e). For **a)** structures *Undoped, Li(24g) v1 and Li(6b)* displacement was $\pm0.09, \pm0.18, \pm0.27$ and $\pm0.36$ Å, for **b)** structure *Li(24g) v2* equilibrium position was $0.58$ Å out-of-plane, with remaining positions being displaced $-0.18, -0.36, +0.36, +0.72$ and $+0.8$ Å.

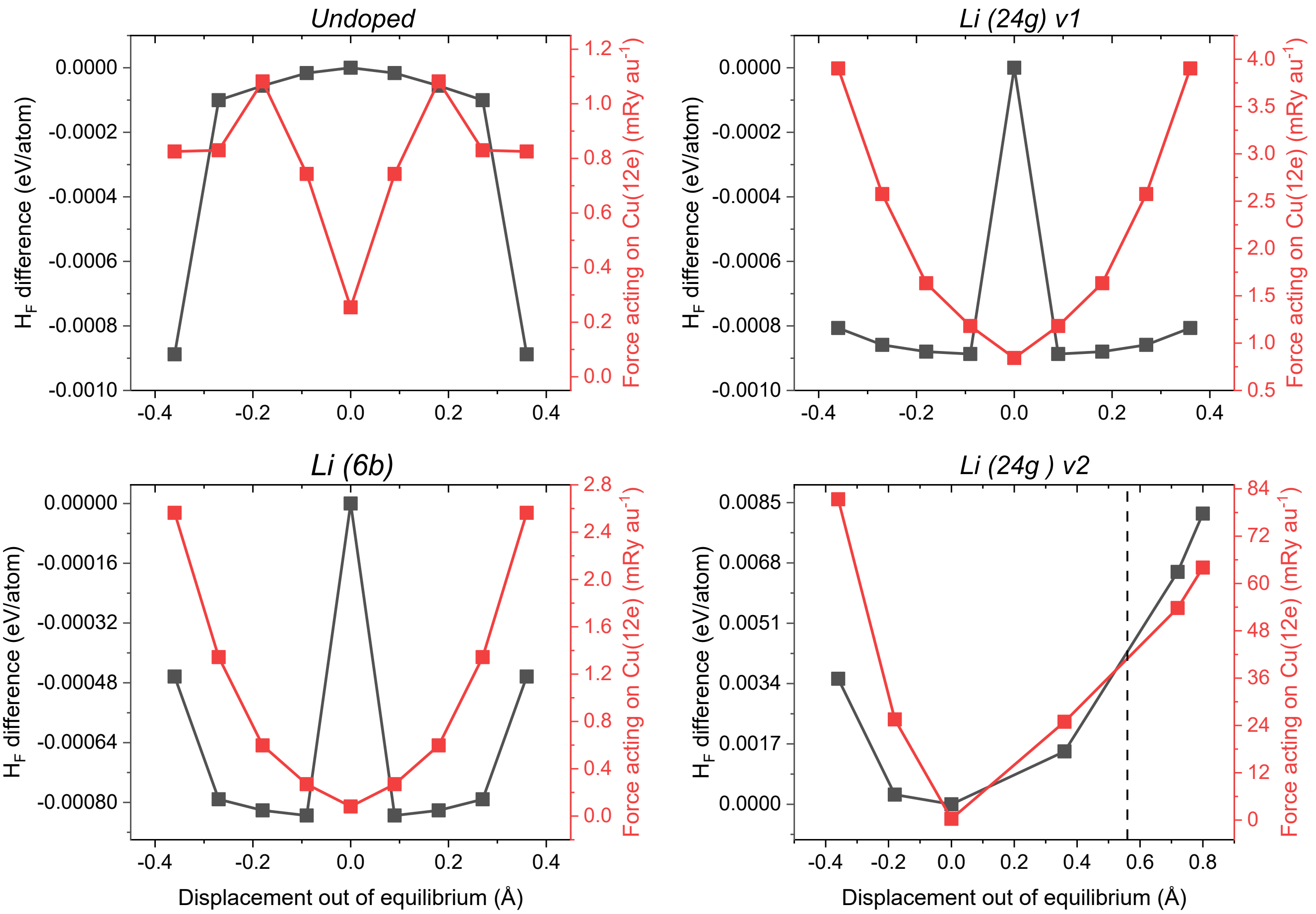

**Figure 7:** Difference in enthalpy of formation $H_F$ and forces acting upon Cu(12e) atom for structures with Cu(12e) displaced out of equilibrium position. Dashed line for *Li(24g) v2* marks position on the $S_3$ plane.

*Li(24g) v1* structure has larger area of the $S_3$ triangle, compared to *Undoped*. Based on results at Figure 7, a small displacement from the equilibrium position results in a lowering of total energy, thus favoring rattling. However, larger displacements increase total energy and should be viewed as a limiting factor for rattling. Forces acting on Cu(12e) are also larger than for *Undoped* structure. These results are consistent with the idea that a larger area of the triangle $S_3$ decreases chemical pressure, thereby reducing the ADP of Cu(12e).
*Li(6b)* structure has roughly the same area $S_{tri}$, however distance to antimony is 0.07 Å longer and Li atom is in proximity to $S_3$ triangle (appendix, Figure 11c). Results are quite similar, with slightly smaller forces and larger increase in energy at 0.36 Å displacement. We interpret those as a consequence of longer distance to antimony, which should result in smaller bond order and less incentive for rattling.
Finally, for *Li(24g) v2* situation is completely different - any displacement from the equilibrium position results in a large increase in energy and forces, indicating that rattling for the out-of-plane Cu(12e) is practically impossible. This result is in agreement previous findings of Umeo et al.[29].

**Table 2**
Bond length **R**, electron density $\rho(r)$, Laplacian of electron density $\nabla^2\rho(r)$, energy density: potential $V[\rho(r)]$, total $H_e[\rho(r)]$, kinetic $G[\rho(r)]$, bond strain index, bond's valency **Val** and multiplicity **Multi** for tetrahedrite structures. All values are for bond critical points (BCP).

| | $x = 0.0$ | R [Å] | $\rho(r)$ [au·$10^{-2}$] | $\nabla^2\rho(r)$ [au·$10^{-2}$] | $V[\rho(r)]$ [au] | $H_e[\rho(r)]$ [au] | $G[\rho(r)]$ [au] | $\lvert V\rvert/G$ [–] | Strain [–] | Val [–] | Multi [–] |
|---|---|---|---|---|---|---|---|---|---|---|---|
| $r_1$ | Cu(12e) – Sb(8c) | 3.227 | 1.779 | 2.903 | -0.084 | -0.038 | 0.046 | 1.841 | 0.004 | 0.000 | 24 |
| $r_1$ | S(2a) – Cu(12e) | 2.269 | 7.560 | 15.824 | -0.920 | -0.440 | 0.480 | 1.918 | 0.001 | 0.333 | 12 |
| $r_1$ | S(24g) – Cu(12d) | 2.266 | 7.867 | 16.986 | -0.865 | -0.411 | 0.454 | 1.906 | 0.000 | 0.333 | 48 |
| $r_1$ | S(24g) – Sb(8c) | 2.477 | 8.006 | 5.711 | -1.048 | -0.571 | 0.531 | 1.973 | 0.001 | 1.000 | 24 |
| $r_1$ | S(24g) – Cu(12e) | 2.230 | 8.430 | 17.656 | -0.946 | -0.451 | 0.495 | 1.911 | 0.001 | 0.333 | 24 |
| | Li 6b $x = 2.0$ | R [Å] | $\rho(r)$ [au·$10^{-2}$] | $\nabla^2\rho(r)$ [au·$10^{-2}$] | $V[\rho(r)]$ [au] | $H_e[\rho(r)]$ [au] | $G[\rho(r)]$ [au] | $\lvert V\rvert/G$ [–] | Strain [–] | Val [–] | Multi [–] |
| $r_1$ | Cu(12e) – Sb(8c) | 3.219 | 1.805 | 2.895 | -0.100 | -0.046 | 0.053 | 1.829 | 0.065 | 0.000 | 24 |
| $r_1$ | S(2a) – Cu(12e) | 2.276 | 7.450 | 15.642 | -0.893 | -0.427 | 0.466 | 1.916 | 0.010 | 0.326 | 12 |
| $r_1$ | S(24g) – Cu(12d) | 2.306 | 7.281 | 16.255 | -0.790 | -0.375 | 0.416 | 1.902 | 0.047 | 0.300 | 48 |
| $r_1$ | S(24g) – Sb(8c) | 2.478 | 7.907 | 6.004 | -1.048 | -0.517 | 0.532 | 1.972 | 0.016 | 0.999 | 24 |
| $r_1$ | S(24g) – Cu(12e) | 2.248 | 8.125 | 17.151 | -0.909 | -0.433 | 0.476 | 1.910 | 0.021 | 0.318 | 24 |
| $r_1$ | S(24g) – Li(6b) | 2.266 | 2.589 | 12.948 | -0.642 | -0.305 | 0.337 | 1.904 | – | 0.250 | 16 |
| | Li 24g $x = 2.0$ | R [Å] | $\rho(r)$ [au·$10^{-2}$] | $\nabla^2\rho(r)$ [au·$10^{-2}$] | $V[\rho(r)]$ [au] | $H_e[\rho(r)]$ [au] | $G[\rho(r)]$ [au] | $\lvert V\rvert/G$ [–] | Strain [–] | Val [–] | Multi [–] |
| $r_1$ | Cu(12e) – Sb(8c) | 3.429 | 2.188 | 2.605 | -0.205 | -0.100 | 0.105 | 1.833 | 0.659 | 0.000 | 24 |
| $r_1$ | S(2a) – Cu(12e) | 2.306 | 7.135 | 14.993 | -0.848 | -0.405 | 0.443 | 1.915 | 0.045 | 0.302 | 12 |
| $r_1$ | S(24g) – Cu(12d) | 2.303 | 7.270 | 16.273 | -0.799 | -0.379 | 0.420 | 1.903 | 0.049 | 0.303 | 48 |
| $r_1$ | S(24g) – Sb(8c) | 2.481 | 7.968 | 4.521 | -1.070 | -0.529 | 0.540 | 1.980 | 0.058 | 1.002 | 24 |
| $r_1$ | S(24g) – Cu(12e) | 2.288 | 7.498 | 16.085 | -0.846 | -0.403 | 0.443 | 1.909 | 0.067 | 0.286 | 24 |
| $r_1$ | S(24g) – Li(24g) | 2.338 | 2.376 | 11.303 | -0.592 | -0.282 | 0.310 | 1.909 | – | 0.333 | 12 |

Regarding topological results, due to the fact that calculations were performed for $P1$ symmetry structures, a large volume of raw data has been obtained. Average values of most important topological parameters for the undoped, Li(6b) $x = 2.0$ and Li(24g) $x = 2.0$ structures are summarized in **Table 2**. The rest of the results can be found in the Appendix.

Calculations predict existence of weak Cu(12e) — Sb(8c) bond (due to presence of bond critical point) for all structures. For structures with Li in 6b structural voids, there are some small, monotonic changes in bond lengths and $\rho(r)$ at the BCP. The biggest ones are for S(24g) — Cu(12d) bond length which increases up to 2.8% and S(24g) — Cu(12e) with 1.83% increase ($x = 3.0$ for both values). Standard deviation for topological parameters for 6b structures is also negligible. This is due to high symmetry of 6b structural void – Li dopant does not introduce a lot of disorder locally. In Li(24g) structures, all bond lengths are longer on average than in Li(6b) structures (except for the S(24g) — Cu(12d) bond). In case of Cu(12e) — Sb(8c) bond, there is great variance in length depending on local presence of the dopant, with the bond being as short as 2.72 Å and as long as 4.24 Å (which is due to Cu(12e) being out-of-plane). This value is reflected in very large strain of Cu(12e) — Sb(8c) bond, 0.659 for $x = 2.0$ structure. Bond length is inversely proportional to $\rho(r)$ and bond strength. This implies that, on average, Li(24g) structures should have weaker bonds and probably lower lattice thermal conductivity.
Globally, values of electron density $\rho(r)$ at the BCP are small, suggesting that tetrahedrite is an ionic system. Usually, small and positive values of Laplacian would also mean that tetrahedrite is mostly ionic, with local depletion of electron density at bond critical points. However, Cremer and Kraka[7] noticed that bond description based on "electrostatical" parameters is not sufficient (eg. there might be covalent $F_2$ — $F_2$ bond with small, positive Laplacian). They started to include energy density values as well, which has been further developed by Espinosa et al.[9] who proposed existence of three different regions based on $V[\rho(r)]$, $H_e[\rho(r)]$ and $G[\rho(r)]$ values. For tetrahedrite, ratio $|V|/G$ in range of $1.804 \div 2.032$ and fulfillment of condition $-G < H_e < 0$ means that structures are in between of closed-shell and shared-shell interaction regions (thought values indicate they are closer to closed-shell, ionic system).

Two other global parameters developed by Mori-Sánchez et al.[20] have been calculated. Flatness of valence electron density $f-index = (\rho_{CCP}(r)^{min})/(\rho_{BCP}(r)^{max}))$ has values of 0.032 ÷ 0.048 which characterized tetrahedrite as very non-metallic. Global charge-transfer index $c$ (average of the ratios of topological charge to nominal oxidation state for each atom) is in range 0.37 ÷ 0.45, which is a typical value for polar compounds (such as III-V crystals).

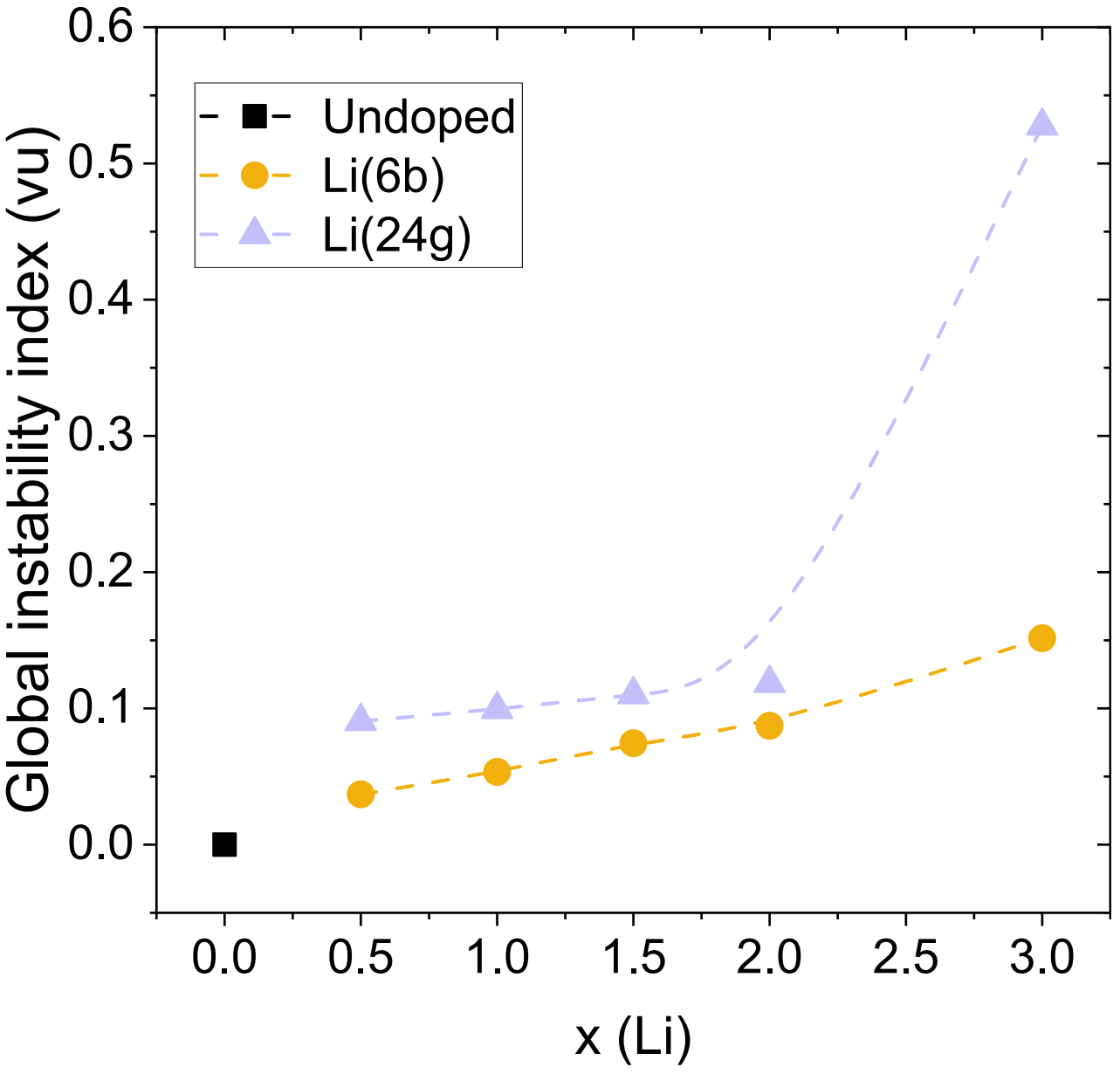

**Figure 8:** Global instability index calculated for undoped and Li-doped tetrahedrite structures. Dashed line is added for improved readability and does not infer actual relation.

Global instability index $G$, calculated based on the difference in ideal and observed valencies, is depicted in **Figure 8**. Typically, for well-defined structures, values of $G$ are around $0.05vu$ (valence units). Larger values indicate some degree of internal strain. There are only a handful of structures for with $G > 0.2vu$ that actually exist in nature. In case of tetrahedrite, only for Li(6b) with $x = 0.5$, value of $G < 0.05vu$. Most of the structures are in range of $0.07 - 0.12vu$, suggesting their experimental synthesis is feasible (consistent with our $H_F$ results). For $x = 3.0$, value of $G$ substantially increases for the Li(24g) structure, reflecting deviations in antimony valencies and bond lengths from ideal values.

For both the electron band structure and density of states, we excluded Li(6b) structures from calculations due to their higher enthalpy of formation, indicating lower energetic favorability. The results presented in Fig. 9-10 are for structures with Li at 24g Wyckoff positions.

The electron band structure for Li-doped tetrahedrite is depicted in **Figure 9** showing the two highest valence bands and two lowest conduction bands. The undoped structure is correctly predicted to be a $p$-type conductor. Introduction of interstitial Li adds one electron to the structure, shifting the Fermi level. For x = 2.0, the material behaves as a classical semiconductor with a band gap decreasing from 0.98 eV (for $x = 0.0$) to 0.49 eV (for $x = 3.0$). Depending on structure, the maximum of the valence band is at H and/or Γ point, while the minimum of the conduction band is also at these points or between them. Increasing Li content above $x = 2.0$ results in the material behaving as an $n$-type conductor, which can be seen for $x = 3.0$.

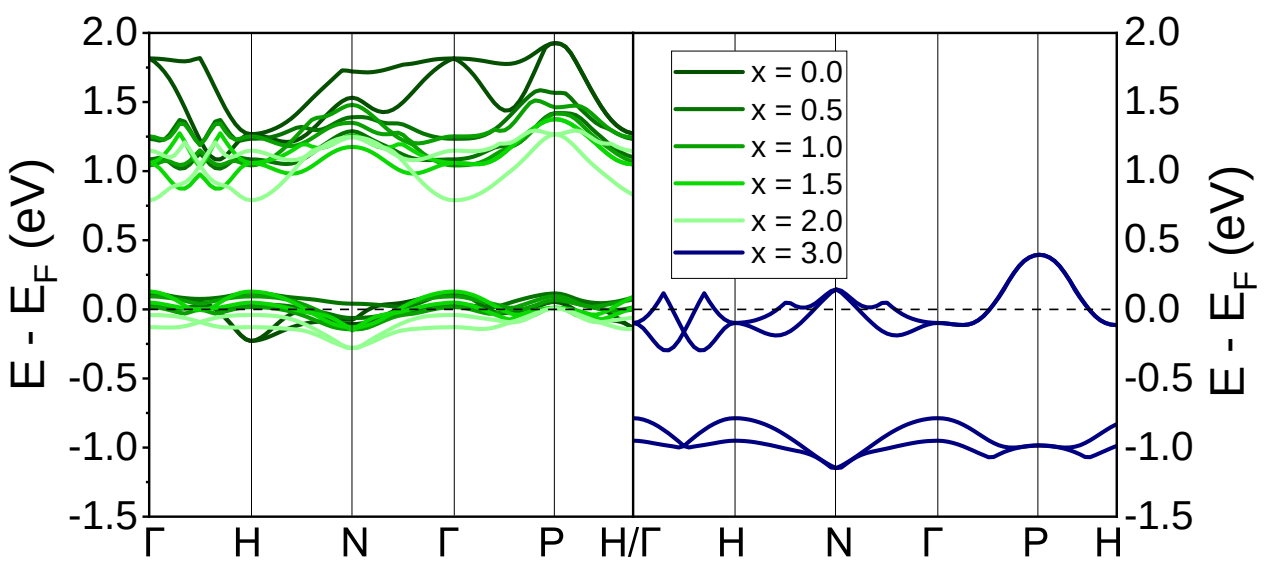

**Figure 9:** Electron band structure for Li-doped tetrahedrite into structural void 24g. For clarity reasons, only two top-most valence and two bottom-most conduction bands are shown.

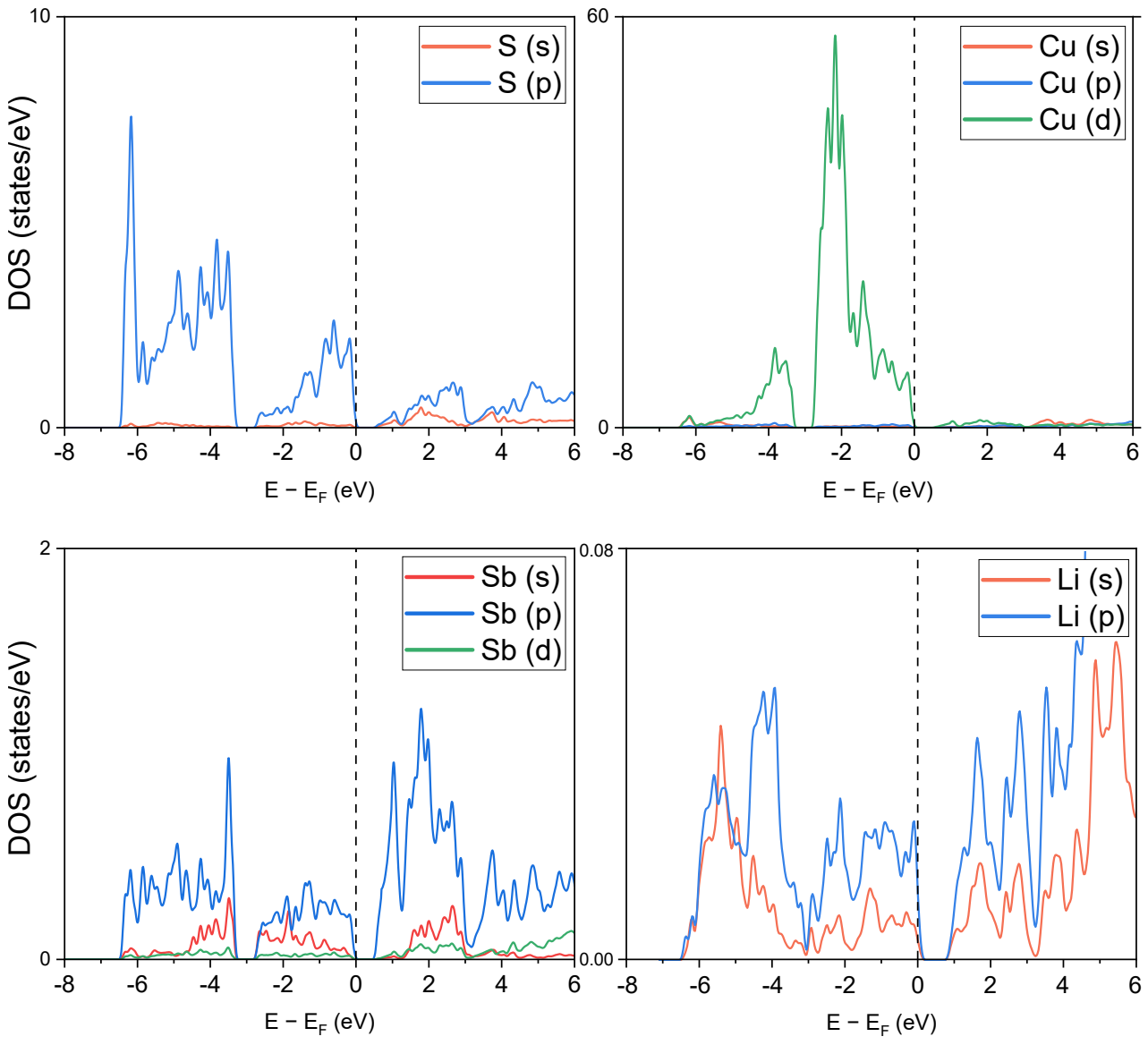

**Figure 10:** Partial Density of States (DOS) for $Li_2(24g)Cu_{12}Sb_4S_{13}$ tetrahedrite. Black dashed line marks Fermi level.

The partial density of states (DOS) for doped tetrahedrite is shown in **Figure 10**. The valence bands closest to the Fermi level are dominated by Cu 3d electrons, with significant contributions from S 3p electrons. These states are crucial, as they define the electronic properties near the Fermi level. Lower energy valence bands are also primarily formed by Cu 3d and S 3p states, indicating the strong hybridization between these orbitals. A notable peak for Cu 3d electrons is observed at -2 eV, which signifies a high density of states in this region. Additionally, there is a prominent peak at -4 eV, consisting of states from both Cu 3d and S 3p electrons. Sb 5p and 5s states are more evenly distributed across the valence band, albeit with lower intensity compared to Cu 3d and S 3p states. This distribution suggests that antimony contributes to the valence band but does not dominate it. The minimum of the conduction band is composed of electron states from Sb, Cu, and S in roughly equal proportions, highlighting the mixed character of these states in the conduction band. As expected, due to its lower concentration, the contribution from Li is much smaller, with minimal impact on the overall electronic structure.

## 4. Conclusions

Based on our findings, the introduction of Li into tetrahedrite's structural voids, both 6b and 24g, appears feasible, although the precise solubility limit is not yet pinpointed. Based on our calculations of the enthalpy of formation ($H_F$), Li-doped tetraehdrite structure is energetically favorable up to $x = 2.0$. Beyond this point, at $x = 3.0$, $H_F$ increases, indicating a less stable state with higher Li content. While acknowledging the speculative nature, fitting a third-order polynomial to the $H_F$ data reveals local minima in range $2.2 \div 2.5$. The global instability index ($G$) also supports the possibility of structures with $x > 2.0$, suggesting that these compositions might be achievable. As calculated electronic band structure demonstrates, achieving more than 2 Li a.p.f.u. would allow for obtaining $n$-type material. Furthermore, the decrease in the average triangle area ($S_{tri}$) for Li(24g) structures should result in increased ADP for Cu(12e) and in turn lower thermal conductivity. Regarding the topological analysis of the charge density, our results indicate that introducing Li into the 6b Wyckoff position does not induce significant local disorder. Most changes are attributed to the increased unit cell parameters. On the other hand, Li(24g) structures have a tendency to change bond lengths and overall are less orderly. $Li_xCu_{12}Sb_4S_{13}$, with $x = 2.1 \div 2.5$ is predicted to be a good $n$-type thermoelectric material with enhanced ADP. However, experimental verification is essential to confirm the feasibility of synthesizing these compositions.

**Acknowledgements**
The research project was supported by the program "Excellence initiative–research university" for the AGH University of Krakow. We also gratefully acknowledge Polish high-performance computing infrastructure PLGrid (HPC Centers: ACK Cyfronet AGH) for providing computer facilities and support within computational grant no. PLG/2022/016040

### Conflict of Interest
The authors declare no conflict of interest.

### Availability Statement
The data that support the findings of this study is available from the corresponding author upon reasonable request.

## CRediT authorship contribution statement

**Krzysztof Kapera:** Conceptualization, Data Curation, Formal analysis, Investigation, Visualization, Writing - Original Draft. **Andrzej Koleżyński:** Conceptualization, Funding acquisition, Project administration, Resources, Supervision, Writing - Review & Editing.

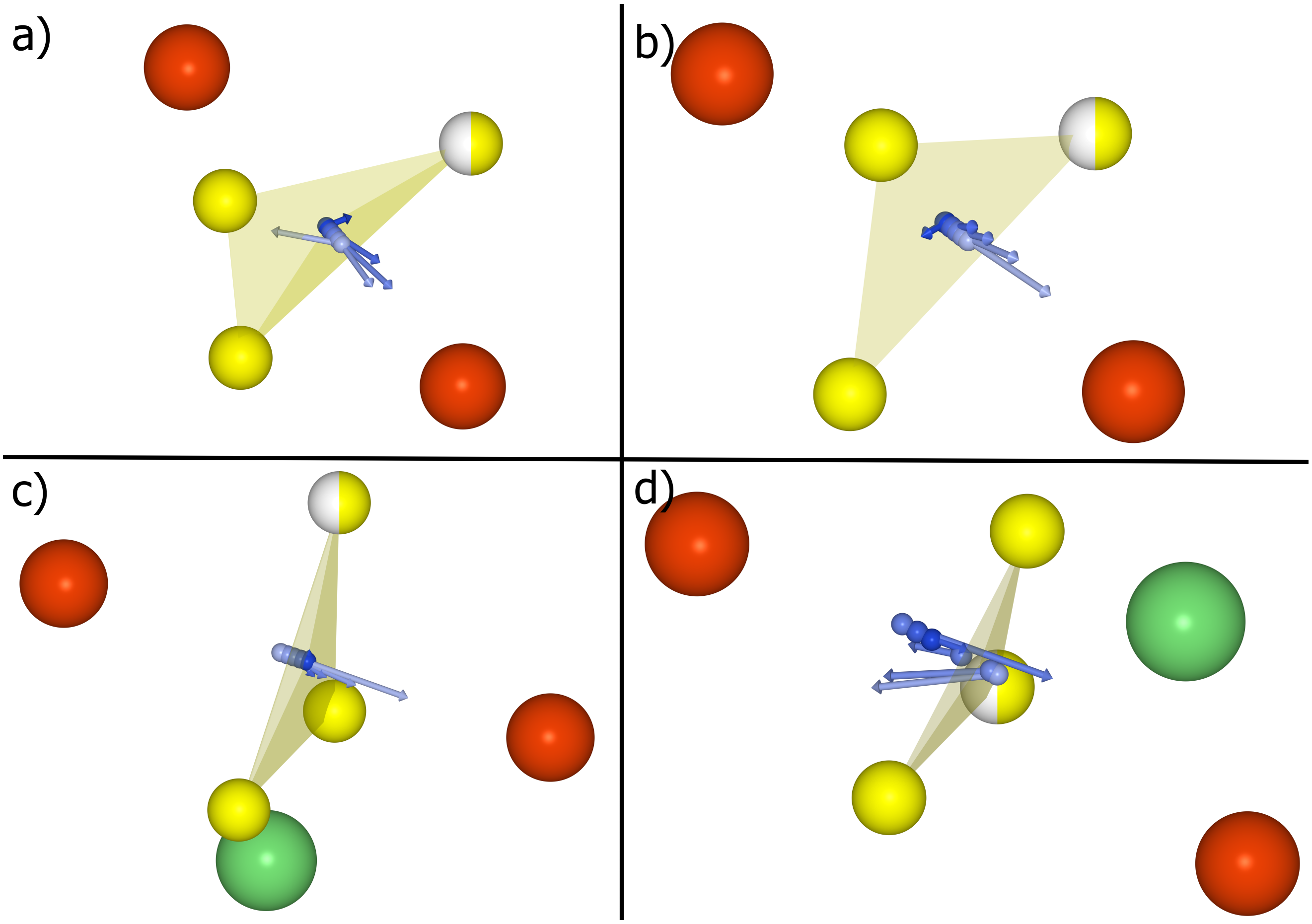

**Figure 11:** Vector forces acting upon Cu(12e) atoms after displacing it out of the equilibrium, for structure **a)** *Undoped*, **b)** *Li(24g) v1*, **c)** *Li(6b)*, **d)** *Li(24g) v2*. Similarly to Figure 6, the darkest hue of blue denotes equilibrium position of Cu(12e).

## A. Vector forces

## B. CRITIC2 tables with topological results

**Table 1**
Bond length **R**, electron density $\boldsymbol{\rho(r)}$, Laplacian of electron density $\boldsymbol{\nabla^2\rho(r)}$, energy density: potential $\boldsymbol{V[\rho(r)]}$, total $\boldsymbol{H_e[\rho(r)]}$, kinetic $\boldsymbol{G[\rho(r)]}$, bond strain index, bond's valency **Val** and multiplicity **Multi** for tetrahedrite structures. All values are for bond critical points (BCP).

| Li 6b $x = 0.5$ | | R [Å] | $\rho(r)$ [au·$10^{-2}$] | $\nabla^2\rho(r)$ [au·$10^{-2}$] | $V[\rho(r)]$ [au] | $H_e[\rho(r)]$ [au] | $G[\rho(r)]$ [au] | $\|V\|/G$ [–] | Strain [–] | Val [–] | Multi [–] |
|---|---|---|---|---|---|---|---|---|---|---|---|
| $r_1$ | Cu(12e) – Sb(8c) | 3.244 | 1.732 | 2.826 | -0.082 | -0.037 | 0.045 | 1.841 | 0.042 | 0.000 | 24 |
| $r_1$ | S(2a) – Cu(12e) | 2.268 | 7.577 | 15.852 | -0.916 | -0.438 | 0.478 | 1.917 | 0.011 | 0.334 | 12 |
| $r_1$ | S(24g) – Cu(12d) | 2.280 | 7.653 | 16.663 | -0.843 | -0.401 | 0.442 | 1.906 | 0.026 | 0.321 | 48 |
| $r_1$ | S(24g) – Sb(8c) | 2.474 | 8.018 | 5.825 | -1.053 | -0.519 | 0.534 | 1.973 | 0.012 | 1.008 | 24 |
| $r_1$ | S(24g) – Cu(12e) | 2.234 | 8.357 | 17.588 | -0.938 | -0.447 | 0.491 | 1.910 | 0.010 | 0.330 | 24 |
| $r_1$ | S(24g) – Li(6b) | 2.267 | 2.570 | 12.862 | -0.650 | -0.309 | 0.341 | 1.906 | – | 0.250 | 4 |
| **Li 6b $x = 1.0$** | | R [Å] | $\rho(r)$ [au·$10^{-2}$] | $\nabla^2\rho(r)$ [au·$10^{-2}$] | $V[\rho(r)]$ [au] | $H_e[\rho(r)]$ [au] | $G[\rho(r)]$ [au] | $\|V\|/G$ [–] | Strain [–] | Val [–] | Multi [–] |
| $r_1$ | Cu(12e) – Sb(8c) | 3.229 | 1.775 | 2.870 | -0.093 | -0.043 | 0.050 | 1.857 | 0.056 | 0.000 | 24 |
| $r_1$ | S(2a) – Cu(12e) | 2.278 | 7.433 | 15.573 | -0.896 | -0.429 | 0.468 | 1.917 | 0.012 | 0.325 | 12 |
| $r_1$ | S(24g) – Cu(12d) | 2.290 | 7.523 | 16.497 | -0.826 | -0.392 | 0.433 | 1.905 | 0.037 | 0.314 | 48 |
| $r_1$ | S(24g) – Sb(8c) | 2.478 | 7.941 | 5.857 | -1.047 | -0.516 | 0.531 | 1.972 | 0.014 | 0.997 | 24 |
| $r_1$ | S(24g) – Cu(12e) | 2.240 | 8.262 | 17.425 | -0.927 | -0.442 | 0.485 | 1.910 | 0.015 | 0.325 | 24 |
| $r_1$ | S(24g) – Li(6b) | 2.269 | 2.564 | 12.814 | -0.646 | -0.307 | 0.339 | 1.906 | – | 0.250 | 8 |
| **Li 6b $x = 1.5$** | | R [Å] | $\rho(r)$ [au·$10^{-2}$] | $\nabla^2\rho(r)$ [au·$10^{-2}$] | $V[\rho(r)]$ [au] | $H_e[\rho(r)]$ [au] | $G[\rho(r)]$ [au] | $\|V\|/G$ [–] | Strain [–] | Val [–] | Multi [–] |
| $r_1$ | Cu(12e) – Sb(8c) | 3.223 | 1.799 | 2.889 | -0.097 | -0.045 | 0.052 | 1.820 | 0.074 | 0.000 | 24 |
| $r_1$ | S(2a) – Cu(12e) | 2.274 | 7.492 | 15.712 | -0.901 | -0.431 | 0.470 | 1.916 | 0.006 | 0.329 | 12 |
| $r_1$ | S(24g) – Cu(12d) | 2.300 | 7.376 | 16.315 | -0.804 | -0.382 | 0.423 | 1.903 | 0.045 | 0.305 | 48 |
| $r_1$ | S(24g) – Sb(8c) | 2.477 | 7.940 | 5.943 | -1.049 | -0.517 | 0.532 | 1.972 | 0.010 | 1.000 | 24 |
| $r_1$ | S(24g) – Cu(12e) | 2.242 | 8.223 | 17.344 | -0.922 | -0.439 | 0.483 | 1.910 | 0.017 | 0.323 | 24 |
| $r_1$ | S(24g) – Li(6b) | 2.261 | 2.608 | 13.054 | -0.651 | -0.309 | 0.342 | 1.905 | – | 0.250 | 12 |
| **Li 6b $x = 2.0$** | | R [Å] | $\rho(r)$ [au·$10^{-2}$] | $\nabla^2\rho(r)$ [au·$10^{-2}$] | $V[\rho(r)]$ [au] | $H_e[\rho(r)]$ [au] | $G[\rho(r)]$ [au] | $\|V\|/G$ [–] | Strain [–] | Val [–] | Multi [–] |
| $r_1$ | Cu(12e) – Sb(8c) | 3.219 | 1.805 | 2.895 | -0.100 | -0.046 | 0.053 | 1.829 | 0.065 | 0.000 | 24 |
| $r_1$ | S(2a) – Cu(12e) | 2.276 | 7.450 | 15.642 | -0.893 | -0.427 | 0.466 | 1.916 | 0.010 | 0.326 | 12 |
| $r_1$ | S(24g) – Cu(12d) | 2.306 | 7.281 | 16.255 | -0.790 | -0.375 | 0.416 | 1.902 | 0.047 | 0.300 | 48 |
| $r_1$ | S(24g) – Sb(8c) | 2.478 | 7.907 | 6.004 | -1.048 | -0.517 | 0.532 | 1.972 | 0.016 | 0.999 | 24 |
| $r_1$ | S(24g) – Cu(12e) | 2.248 | 8.125 | 17.151 | -0.909 | -0.433 | 0.476 | 1.910 | 0.021 | 0.318 | 24 |
| $r_1$ | S(24g) – Li(6b) | 2.266 | 2.589 | 12.948 | -0.642 | -0.305 | 0.337 | 1.904 | – | 0.250 | 16 |
| **Li 6b $x = 3.0$** | | R [Å] | $\rho(r)$ [au·$10^{-2}$] | $\nabla^2\rho(r)$ [au·$10^{-2}$] | $V[\rho(r)]$ [au] | $H_e[\rho(r)]$ [au] | $G[\rho(r)]$ [au] | $\|V\|/G$ [–] | Strain [–] | Val [–] | Multi [–] |
| $r_1$ | Cu(12e) – Sb(8c) | 3.167 | 1.970 | 2.836 | -0.115 | -0.054 | 0.061 | 1.884 | 0.060 | 0.000 | 24 |
| $r_1$ | S(2a) – Cu(12e) | 2.307 | 7.025 | 14.666 | -0.845 | -0.404 | 0.441 | 1.917 | 0.039 | 0.300 | 12 |
| $r_1$ | S(24g) – Cu(12d) | 2.329 | 6.975 | 15.465 | -0.760 | -0.361 | 0.399 | 1.903 | 0.063 | 0.281 | 48 |
| $r_1$ | S(24g) – Sb(8c) | 2.516 | 7.276 | 6.364 | -1.009 | -0.499 | 0.513 | 1.969 | 0.039 | 0.899 | 24 |
| $r_1$ | S(24g) – Cu(12e) | 2.271 | 7.764 | 16.347 | -0.876 | -0.417 | 0.458 | 1.911 | 0.041 | 0.299 | 24 |
| $r_1$ | S(24g) – Li(6b) | 2.284 | 2.501 | 12.431 | -0.626 | -0.297 | 0.329 | 1.905 | – | 0.250 | 20 |

**Table 2**
Bond length **R**, electron density $\boldsymbol{\rho(r)}$, Laplacian of electron density $\boldsymbol{\nabla^2\rho(r)}$, energy density: potential $\boldsymbol{V[\rho(r)]}$, total $\boldsymbol{H_e[\rho(r)]}$, kinetic $\boldsymbol{G[\rho(r)]}$, bond strain index, bond's valency **Val** and multiplicity **Multi** for tetrahedrite structures. All values are for bond critical points (BCP).

| Li 24g $x = 0.5$ | | R [Å] | $\rho(r)$ [au·$10^{-2}$] | $\nabla^2\rho(r)$ [au·$10^{-2}$] | $V[\rho(r)]$ [au] | $H_e[\rho(r)]$ [au] | $G[\rho(r)]$ [au] | $\lvert V\rvert/G$ [–] | Strain [–] | Val [–] | Multi [–] |
|---|---|---|---|---|---|---|---|---|---|---|---|
| $r_1$ | Cu(12e) – Sb(8c) | 3.234 | 2.302 | 2.936 | -0.222 | -0.107 | 0.115 | 1.841 | 0.396 | 0.000 | 24 |
| $r_1$ | S(2a) – Cu(12e) | 2.285 | 7.393 | 15.450 | -0.886 | -0.424 | 0.462 | 1.917 | 0.047 | 0.321 | 12 |
| $r_1$ | S(24g) – Cu(12d) | 2.278 | 7.672 | 16.717 | -0.838 | -0.398 | 0.440 | 1.906 | 0.022 | 0.323 | 48 |
| $r_1$ | S(24g) – Sb(8c) | 2.494 | 7.815 | 4.776 | -1.010 | -0.499 | 0.511 | 1.977 | 0.037 | 0.958 | 24 |
| $r_1$ | S(24g) – Cu(12e) | 2.258 | 7.982 | 16.815 | -0.894 | -0.426 | 0.468 | 1.910 | 0.045 | 0.310 | 24 |
| $r_1$ | S(24g) – Li(6b) | 2.391 | 2.172 | 10.194 | -0.552 | -0.263 | 0.289 | 1.913 | – | 0.250 | 3 |
| $r_1$ | Sb(8c) – Li(24g) | 2.720 | 1.455 | 5.771 | -0.451 | -0.218 | 0.232 | 1.938 | – | 0.250 | 1 |
| **Li 24g $x = 1.0$** | | R [Å] | $\rho(r)$ [au·$10^{-2}$] | $\nabla^2\rho(r)$ [au·$10^{-2}$] | $V[\rho(r)]$ [au] | $H_e[\rho(r)]$ [au] | $G[\rho(r)]$ [au] | $\lvert V\rvert/G$ [–] | Strain [–] | Val [–] | Multi [–] |
| $r_1$ | Cu(12e) – Sb(8c) | 3.216 | 2.684 | 3.148 | -0.355 | -0.174 | 0.182 | 1.937 | 0.578 | 0.000 | 24 |
| $r_1$ | S(2a) – Cu(12e) | 2.298 | 7.269 | 15.220 | -0.859 | -0.410 | 0.448 | 1.915 | 0.052 | 0.310 | 12 |
| $r_1$ | S(24g) – Cu(12d) | 2.285 | 7.552 | 16.644 | -0.834 | -0.396 | 0.438 | 1.905 | 0.033 | 0.317 | 48 |
| $r_1$ | S(24g) – Sb(8c) | 2.477 | 8.030 | 5.324 | -1.041 | -0.514 | 0.527 | 1.975 | 0.049 | 1.008 | 24 |
| $r_1$ | S(24g) – Cu(12e) | 2.273 | 7.747 | 16.449 | -0.865 | -0.412 | 0.453 | 1.909 | 0.057 | 0.298 | 24 |
| $r_1$ | S(24g) – Li(24g) | 2.337 | 2.394 | 11.359 | -0.452 | -0.212 | 0.240 | 1.804 | – | 0.333 | 6 |
| **Li 24g $x = 1.5$** | | R [Å] | $\rho(r)$ [au·$10^{-2}$] | $\nabla^2\rho(r)$ [au·$10^{-2}$] | $V[\rho(r)]$ [au] | $H_e[\rho(r)]$ [au] | $G[\rho(r)]$ [au] | $\lvert V\rvert/G$ [–] | Strain [–] | Val [–] | Multi [–] |
| $r_1$ | Cu(12e) – Sb(8c) | 3.265 | 2.559 | 2.920 | -0.302 | -0.147 | 0.155 | 1.926 | 0.562 | 0.000 | 24 |
| $r_1$ | S(2a) – Cu(12e) | 2.301 | 7.212 | 15.148 | -0.856 | -0.409 | 0.447 | 1.915 | 0.046 | 0.306 | 12 |
| $r_1$ | S(24g) – Cu(12d) | 2.295 | 7.394 | 16.432 | -0.812 | -0.385 | 0.426 | 1.904 | 0.041 | 0.309 | 48 |
| $r_1$ | S(24g) – Sb(8c) | 2.475 | 8.133 | 4.954 | -1.056 | -0.522 | 0.534 | 1.977 | 0.052 | 1.015 | 24 |
| $r_1$ | S(24g) – Cu(12e) | 2.281 | 7.612 | 16.264 | -0.857 | -0.408 | 0.449 | 1.909 | 0.062 | 0.298 | 24 |
| $r_1$ | S(24g) – Li(24g) | 2.345 | 2.356 | 11.178 | -0.588 | -0.280 | 0.308 | 1.910 | – | 0.333 | 9 |
| **Li 24g $x = 2.0$** | | R [Å] | $\rho(r)$ [au·$10^{-2}$] | $\nabla^2\rho(r)$ [au·$10^{-2}$] | $V[\rho(r)]$ [au] | $H_e[\rho(r)]$ [au] | $G[\rho(r)]$ [au] | $\lvert V\rvert/G$ [–] | Strain [–] | Val [–] | Multi [–] |
| $r_1$ | Cu(12e) – Sb(8c) | 3.429 | 2.188 | 2.605 | -0.205 | -0.100 | 0.105 | 1.833 | 0.659 | 0.000 | 24 |
| $r_1$ | S(2a) – Cu(12e) | 2.306 | 7.135 | 14.993 | -0.848 | -0.405 | 0.443 | 1.915 | 0.045 | 0.302 | 12 |
| $r_1$ | S(24g) – Cu(12d) | 2.303 | 7.270 | 16.273 | -0.799 | -0.379 | 0.420 | 1.903 | 0.049 | 0.303 | 48 |
| $r_1$ | S(24g) – Sb(8c) | 2.481 | 7.968 | 4.521 | -1.070 | -0.529 | 0.540 | 1.980 | 0.058 | 1.002 | 24 |
| $r_1$ | S(24g) – Cu(12e) | 2.288 | 7.498 | 16.085 | -0.846 | -0.403 | 0.443 | 1.909 | 0.067 | 0.286 | 24 |
| $r_1$ | S(24g) – Li(24g) | 2.338 | 2.376 | 11.303 | -0.592 | -0.282 | 0.310 | 1.909 | – | 0.333 | 12 |
| **Li 24g $x = 3.0$** | | R [Å] | $\rho(r)$ [au·$10^{-2}$] | $\nabla^2\rho(r)$ [au·$10^{-2}$] | $V[\rho(r)]$ [au] | $H_e[\rho(r)]$ [au] | $G[\rho(r)]$ [au] | $\lvert V\rvert/G$ [–] | Strain [–] | Val [–] | Multi [–] |
| $r_1$ | Cu(12e) – Sb(8c) | 3.460 | 1.978 | 2.411 | -0.076 | -0.035 | 0.041 | 1.936 | 0.650 | 0.000 | 24 |
| $r_1$ | S(2a) – Cu(12e) | 2.391 | 6.203 | 12.330 | -0.724 | -0.346 | 0.377 | 1.918 | 0.183 | 0.254 | 12 |
| $r_1$ | S(24g) – Cu(12d) | 2.307 | 7.221 | 16.101 | -0.804 | -0.382 | 0.422 | 1.905 | 0.052 | 0.299 | 48 |
| $r_1$ | S(24g) – Sb(8c) | 3.083 | 5.032 | 3.256 | -0.536 | -0.264 | 0.272 | 2.032 | 0.949 | 0.590 | 24 |
| $r_1$ | S(24g) – Cu(12e) | 2.307 | 7.256 | 15.423 | -0.825 | -0.393 | 0.432 | 1.911 | 0.150 | 0.269 | 24 |
| $r_1$ | S(24g) – Li(24g) | 2.360 | 2.399 | 11.257 | -0.590 | -0.281 | 0.309 | 1.910 | – | 0.333 | 12 |

**Declaration of interests**

☒The authors declare that they have no known competing financial interests or personal relationships that could have appeared to influence the work reported in this paper.

☐The authors declare the following financial interests/personal relationships which may be considered as potential competing interests:

## CRediT authorship contribution statement

**Krzysztof Kapera:** Conceptualization, Data Curation, Formal analysis, Investigation, Visualization, Writing - Original Draft,

**Andrzej Koleżyński:** Conceptualization, Funding acquisition, Project administration, Resources, Supervision, Writing - Review & Editing